\documentclass[showpacs,twocolumn,superscriptaddress,nofootinbib]{revtex4-1} 

\usepackage[utf8]{inputenc}
\usepackage{hyperref}
\hypersetup{colorlinks=true,linkcolor=blue,citecolor=cyan}
\usepackage{graphicx}
\usepackage{xcolor, soul}
\sethlcolor{green}
\usepackage{dcolumn}
\usepackage{bm}
\usepackage{color}
\usepackage{enumitem}
\usepackage{mathrsfs}
\usepackage{amsmath}
\usepackage{amssymb}
\usepackage{cleveref}

\usepackage{soul,xcolor}
\setstcolor{red}

\begin{document}

\title{Extending the Comisso–Asenjo Energy Extraction Mechanism to Pure Lovelock Gravity}

\author{Chen Zhou}
\email{chenzhou@hit.edu.cn}
\affiliation{School of Physics, Harbin Institute of Technology, Harbin 150001, People’s Republic of China}

\author{Ikhtiyor Eshtursunov}
\email{eshtursunovikhtiyor@gmail.com}
\affiliation{New Uzbekistan University, Movarounnahr str. 1, Tashkent 100000, Uzbekistan}

\author{Sanjar Shaymatov}
\email{sanjar@astrin.uz}
\affiliation{Institute of Fundamental and Applied Research, National Research University TIIAME, Kori Niyoziy 39, Tashkent 100000, Uzbekistan}
\affiliation{Tashkent University of Applied Sciences, Gavhar Str. 1, Tashkent 100149, Uzbekistan}
\affiliation{National University of Uzbekistan, Tashkent 100174, Uzbekistan}

\author{Chengxun Yuan}
   \email{yuancx@hit.edu.cn}
\affiliation{School of Physics, Harbin Institute of Technology, Harbin 150001, People’s Republic of China}

\date{\today}
\begin{abstract}

In this paper, we extend the Comisso–Asenjo magnetic reconnection (MR) mechanism to rotating black holes (BHs) in pure Lovelock/Gauss--Bonnet (GB) gravity in dimension $2N+2\leq D\leq 4N+1$ (where $N$ is the Lovelock polynomial degree of $N$th order term in the action). We perform a comprehensive analysis of the efficiency and power of extracted energy by exploring the effects of the spin parameter, plasma magnetization, magnetic field orientation, and reconnection location. Our results reveal distinctive energetic features of pure Lovelock BHs relative to their Einstein counterparts, except in the special case of $D=3N+1$, where the two theories coincide. Our results demonstrate that magnetic reconnection becomes increasingly efficient in extracting rotational energy from rapidly rotating pure Lovelock BHs with single rotation configuration as the spacetime dimension increases from $D=6$ to $9$. Furthermore, the Comisso--Asenjo MR mechanism can produce higher extraction power than the Blandford--Znajek (BZ) process in certain regions of parameter space because of its enhanced energy extraction rate. These results show that magnetic reconnection is an efficient mechanism for extracting rotational energy from rapidly rotating pure Lovelock/GB BHs, highlighting their relevance to high-energy astrophysical phenomena.

\end{abstract}
\maketitle

\section{Introduction}
\label{introduction}

Lovelock gravity~\cite{Lovelock71} represents the unique higher-dimensional generalization of Einstein gravity that preserves second-order field equations. Its action is constructed from homogeneous polynomials of degree $N$ in the Riemann tensor, with $N=1$ corresponding to the Einstein--Hilbert term, $N=2$ to the Gauss--Bonnet term, and higher orders representing successive curvature corrections. The Lovelock Lagrangian is given by a sum of dimensionally extended Euler densities, each associated with an independent coupling constant. A distinguishing feature of the theory is that, despite the presence of higher-order curvature terms, the field equations remain second order in the metric, thereby avoiding ghost degrees of freedom. Higher-curvature corrections and extra dimensions arise naturally in string theory, where the Gauss--Bonnet (GB) term appears as the leading one-loop correction to the low-energy effective action~\cite{Zwiebach85,Sen05}. This provides a strong theoretical motivation for treating GB and, more generally, Lovelock gravity as string-inspired extensions of Einstein gravity. The first static vacuum BH solution in Einstein--Gauss--Bonnet gravity was obtained by Boulware and Deser~\cite{Boulware85}. Although exact solutions become increasingly difficult to derive at higher Lovelock orders $N$, static BH solutions have subsequently been constructed for arbitrary order~\cite{Wheeler86a,Wheeler86b,Whitt88}.

Black strings and black branes represent natural higher-dimensional extensions of BHs, characterized by event horizons stretched along one or more extra spatial dimensions. In Einstein gravity, they can be constructed by trivially extending known BH geometries with Ricci-flat directions. Such a prescription is generally not applicable in Lovelock gravity. Instead, extended objects arise in pure Lovelock gravity, where the gravitational action is composed of a single Lovelock term of order $N$~\cite{Kastor06,Giribet06}. Another important family of solutions is provided by dimensionally continued BHs, obtained by fixing the Lovelock coupling constants in terms of a unique vacuum cosmological constant $\Lambda$~\cite{BTZ1994,CTZ2000}. Their pure Lovelock BHs and associated thermodynamic and geodesic behavior have been analyzed in detail in Refs.~\cite{Cai06,Dadhich2022PDU}.

These developments established a strong theoretical foundation for pure Lovelock gravity, which was formulated on a rigorous geometric basis in Ref.~\cite{Dadhich12}. In this framework, a Lovelock generalization of the Riemann tensor was introduced~\cite{Camanho16}, whose Bianchi identity gives rise to a divergence-free symmetric rank-two tensor analogous to the Einstein tensor. This construction preserves the fundamental geometric structure of General Relativity while extending it to higher-order curvature theories. A key property of Einstein gravity is its kinematic character in three dimensions. Since the Riemann tensor is completely determined by the Ricci tensor—or equivalently, the Weyl tensor vanishes identically—three-dimensional vacuum spacetimes admit no local gravitational degrees of freedom and therefore no nontrivial vacuum solutions. This property forms the basis for the notion of kinematic gravity. 

A distinctive feature of pure Lovelock gravity is that the kinematic character of gravity extends universally to all critical odd dimensions, $D=2N+1$~\cite{Dadhich12,Camanho16}. In these dimensions, the Lovelock Riemann tensor is completely determined by the corresponding Lovelock Ricci tensor, implying the absence of nontrivial vacuum solutions and the identically vanishing Lovelock Weyl tensor. Moreover, pure Lovelock gravity uniquely admits stable bound orbits around static BHs in higher dimensions, a property absent in other Lovelock theories~\cite{Dadhich13}. These remarkable features distinguish pure Lovelock gravity from other higher-curvature theories. Another noteworthy property of pure Lovelock gravity is its close correspondence with four-dimensional vacuum Einstein gravity in $D=3N+1$ dimensions~\cite{Chakraborty18,Gannouji19}. This correspondence has motivated extensive studies of compact stellar configurations, including the derivation of compactness limits for stable stars in dimensions $D\geq 2N+1$~\cite{Dadhich14,Dadhich16a,Chakraborty20egb,Dadhich17,Dadhich17b}. In parallel, the Brown--York quasi-local energy of pure Lovelock BHs~\cite{Chakraborty15} and the contribution of gravitational field energy to the Buchdahl compactness bound for static fluid spheres~\cite{Dadhich20:JCAP} have also been investigated. It was also shown that a non-extremal Buchdahl star cannot be extremalized by test particle accretion \cite{Shaymatov23PLB}. An intriguing property of pure Lovelock BHs is that they always obey the weak cosmic censorship conjecture in all dimensions $D>4N+1$ \cite{Shaymatov2022JCAP...10..060S}. 

In this work, we investigate the energetics of rotating BHs in pure Lovelock gravity by extending the magnetic reconnection mechanism proposed by Comisso and Asenjo \cite{Comisso21}. We explore how the efficiency and power of rotational energy extraction depend on the BH spin, plasma magnetization, magnetic field orientation, and the location of the reconnection region. Our analysis reveals characteristic features that distinguish pure Lovelock BHs from their four-dimensional vacuum Einstein counterparts, except in the special case of $D=3N+1$, where the two theories exhibit close correspondence. 

Astrophysical BHs are widely believed to power some of the most energetic phenomena in the Universe, including active galactic nuclei, X-ray binaries, and gamma-ray bursts~\cite{King01ApJ,Peterson:97book,Meszaros06}. To this end, several efficient mechanisms have been proposed to extract rotational energy from astrophysical BHs, including the Blandford--Znajek \cite{Blandford1977}, magnetic Penrose \cite{Penrose:1969pc,Bhat85,Parthasarathy86ApJ}, and more recently the magnetic reconnection mechanisms \cite{Koide08ApJ,Parfrey19PRL}. These processes are believed to power some of the most energetic astrophysical phenomena \cite{Bhat85,Parthasarathy86ApJ,Blandford1977,Rees:1984si,Peterson:97book,Meszaros06,King01ApJ,McKinney07}, such as relativistic jets and high-energy particle acceleration~\cite{Blandford1977,Wagh89}. Although these mechanisms extract rotational energy from rotating BHs, they rely on distinct physical processes. The Blandford--Znajek mechanism operates through magnetohydrodynamic interactions in a force-free magnetosphere, whereas the magnetic Penrose process exploits the combined effects of frame dragging and electromagnetic interactions, yielding energy extraction efficiencies that can exceed 100\% for charged particles \cite{Dhurandhar1984PRD.30.1625,Wagh85ApJ,Abdujabbarov11,Tursunov:2019oiq,Shaymatov24PRD.110d4042S,Xamidov24EPJC,Dadhich18MNRAS,Shaymatov24EPJC,Shaymatov22b,Khamidov25JCAP...03..053X}. 

Unlike the Blandford--Znajek and magnetic Penrose mechanisms, magnetic reconnection occurs when frame dragging near the BH horizon creates antiparallel magnetic field configurations. Operating within the ergosphere, it enables negative-energy particles to fall into the BH while accelerating the remaining particles to higher energies, thereby extracting rotational energy. Although slow MR is inefficient for powering the most energetic astrophysical sources \cite{Koide08ApJ}, relativistic MR can efficiently accelerate particles and drive energetic outflows \cite{Parfrey19PRL}. Comisso and Asenjo, who showed that the extraction efficiency and power strongly depend on the spin of the BH, developed a quantitative framework for MR-driven energy extraction \cite{Comisso21}. This mechanism has since been extensively investigated and compared with other energy extraction processes in rapidly rotating BHs \cite{Liu22ApJ,Wei22,Carleo22,Khodadi22,Wang22,Khodadi23MR_JCAP,Shaymatov24MR,Zhang24JCAP...07..042Z,Chen24PRD.110f3003C,Shen25PRD.111b3003S,Long25EPJC...85...26L,Rodriguez25PDU,YuChih25PRD.112j4016Y,Wang25JCAP,jaguri2026,Zeng25PRD.112f4032Z,Zeng25PRD.112f4080Z,eshtursunov2026b,Cheng25EPJC...85.1130C,Eshtursunov:2026kgr}. Together with future observations, these mechanisms will help establish a more comprehensive understanding of the extreme environments surrounding BHs. 

This paper is organized as follows: In Sec.~\ref{Sec:GB}, we briefly introduce pure GB rotating BH spacetime and about motivation for pure Lovelock theory. In Sec.~\ref{Sec:GB_MR}, we follow the Comisso–Asenjo mechanism to explore energy extraction driven from rapidly rotating BHs in pure Lovelock gravity. We devote ourselves to analyze how the efficiency, power of rotational energy extraction and the overall energy extraction rate depend on the BH spin, plasma magnetization, magnetic field orientation, and the location of the reconnection region for allowed phase-space regions for accelerated and decelerated plasma energies in Sec.~\ref{Sec:MP-Power-MR}. Finally, we end up with our findings and conclusions in Sec.~\ref{Sec:con}. Throughout the paper, we use a system of geometric units with $G_{\rm{N}}=c=1$.

\section{Pure Lovelock/GB rotating black hole spacetime}\label{Sec:GB}

In contrast to Einstein gravity, where the Myers--Perry metric provides an exact higher-dimensional rotating vacuum BH solution \cite{Myers-Perry86} with distinctive features \cite{An18,Shaymatov19a,Prabhu10,Shaymatov20a,Shaymatov20b,2022IJMPD..3150120D,2021PDU....3100758S}, pure Lovelock/GB gravity admits no exact rotating vacuum solution. However, a rotating pure GB BH metric was proposed in Ref.~\cite{Dadhich-Ghosh13} using the method introduced in Ref.~\cite{Dadhich13b}, which derives rotating spacetimes without explicitly solving the field equations. The method is based on two physically motivated conditions: photons moving along the rotation axis experience no three-acceleration, while timelike particles reproduce the appropriate Newtonian acceleration. Applied to an ellipsoidal geometry in six dimensions, with the Newtonian acceleration replaced by its pure GB counterpart, this method yields a metric exhibiting the essential features of a rotating BH. Although it does not exactly satisfy the pure GB vacuum field equations, it remains valid to leading order.

The metric describing the Myers-Perry higher dimensional rotating BH \cite{Myers-Perry86,Dadhich13GB} is given by 
\begin{eqnarray}\label{D}
ds^2&=&-dt^2+r^2d\beta^2 + (r^2+a^2_{n})\left(d\mu_{n}^2+\mu_{n}^2d\phi^2_{n}\right)\nonumber\\&+&\frac{\mu r}{\Pi F}\left(dt +a_{n}\mu_{n}^2d\phi_{i}\right) +\frac{\Pi F}{\Delta}dr^2\, ,
\end{eqnarray}
where
\begin{eqnarray}
F &=& 1-\frac{a_{n}^2\mu_{n}^2} {r^2+a_{n}^2}\, , \nonumber\\
\Pi &=&(r^2+a_1^2)...(r^2+a_n^2) \, ,
\end{eqnarray}
with 
\begin{eqnarray}\label{Eq:delta1}
\Delta = \frac{\left(r^2+a^2\right)...\left(r^2+a_{n}^2\right)}{r^{2n-2}} -2\mu r^{2n+3-D}\, .
\end{eqnarray}
We note that $n=[(D-1)/2]$ refers to the maximum number of rotation parameters in the given dimension $D$, while $\mu$ and $a_n$ are mass and rotation parameters, respectively. Here $\mu_n$ and $\beta$ has a relation for odd and even $D$ dimensions and are, respectively, given by 
\begin{eqnarray}
\Sigma \mu_n^2 + \beta^2 &=& 1\, ,\\
\Sigma \mu_n^2 &=& 1\, .
\end{eqnarray}

It should be noted that the effective metric in pure Lovelock gravity is defined by arbitrary order $N$, which is the Lovelock polynomial degree in action with only one $N$th order term, i.e., $\mu/r^\alpha$ with $\alpha=(D-2N-1)/N$ in dimensions $D>2N+1$~\cite{Dadhich11GB,Dadhich-Ghosh13}. Interestingly, pure Lovelock gravity is kinematic, leading to the vacuum solution, which is trivial with a vanishing Lovelock Riemann tensor in odd dimension $D=2N+1$ (see, e.g., \cite{Dadhich12,Dadhich16b}). As a result, in the last term of Eq.~(\ref{Eq:delta1}), the power of $r$ can be defined by ${2n-\alpha}$. Eq.~(\ref{Eq:delta1}) then produces as:
\begin{eqnarray}\label{Eq:delta2}
\Delta = \frac{\left(r^2+a^2\right)...\left(r^2+a_{n}^2\right)}{r^{2n-2}} -2\mu r^{2n-\alpha}\, .
\end{eqnarray}
In this way, we can obtain an effective metric for the pure Lovelock analogue of the Myers–Perry rotating BH. While it is not an exact solution of the pure Lovelock vacuum field equations, it satisfies them at leading order and exhibits all the essential features of a rotating BH. $\Delta=0$ solves to give the BH horizon with real positive roots 
\begin{eqnarray}\label{Eq:delta3}
(r^2+a_1^2)...(r^2+a_n^2) - 2\mu\,r^{\left(2n-\frac{D-2N-1}{N}\right)} =0\, .
\end{eqnarray}

From the above polynomial equation, the following condition must be satisfied for any Lovelock order:
\begin{eqnarray}\label{Eq:con1}
2N+2\leq D\leq 2N(n+1)+1\, .
\end{eqnarray} 
This condition determines the allowed dimensional window. For the Lovelock order $N=2$, the corresponding range is $6\leq D \leq9$ for $n=1$, while it becomes $6 \leq D \leq 13$ for $n=2$. For further analysis, we consider the BH with a single rotation parameter, $n=1$, and restrict the allowed dimensional window to $6\leq D \leq9$.

\section{The Comisso–Asenjo MR Mechanism for Energy Extraction}\label{Sec:GB_MR}

Testing energetic processes around higher dimensional BHs can be very important because it provides new concepts about complex relations of spacetime geometries, gravitational dynamics, and high-energy phenomena in higher dimensions. Such studies are significantly important because, in contrast to Einstein theory in four dimensions, the horizon structure, ergosphere, and the geodesic motion of test particles can be changed in higher dimensions. Therefore, it has value to extend the Comisso and Asenjo magnetic reconnection mechanism to pure Lovelock/GB BHs. It must be emphasized that pure Lovelock theory in $D = 3N +1$  dimensions for $n=1$ is similar to vacuum Einstein gravity in $D=4$ dimensions. We further employ the Comisso and Asenjo MR mechanism to investigate rotating BHs with a single rotation configuration in dimensions $6\leq D \leq9$ in pure Lovelock gravity. To understand how the energy extraction process occurs, we now investigate the behavior of plasma energy density under the influence of strong frame dragging effects within the ergospheres of the higher-dimensional BHs in pure Lovelock gravity. We begin the analysis by introducing the Minkowski-like locally nonrotating reference frame, i.e., the zero angular momentum observer (ZAMO), in this frame we examine plasma energetics. In the ZAMO frame, a spacetime line element can be written as
\begin{figure} 
\begin{center} \begin{tabular}{c c} 
\includegraphics[scale=0.45]{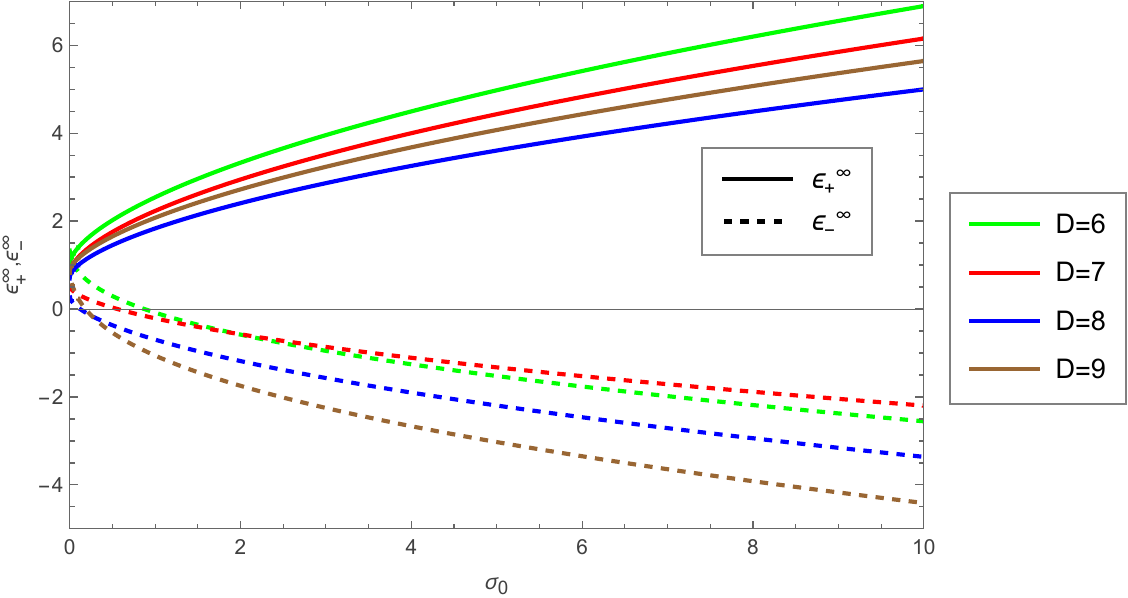}\hspace{1cm} 
\end{tabular} \caption{\label{Fig:energyatinf} {The relationship between the magnetization parameter and the accelerated/decelerated plasma energies at infinity per unit enthalpy for rotating pure Lovelock Bhs in $D=6,7,8,9$ dimensions with a single rotation when their rotation parameters approach their maximum values.}} 
\end{center}
\end{figure}
\begin{figure*}[ht]
\centering
\begin{tabular}{cc}
  \includegraphics[width=0.48\textwidth]
  {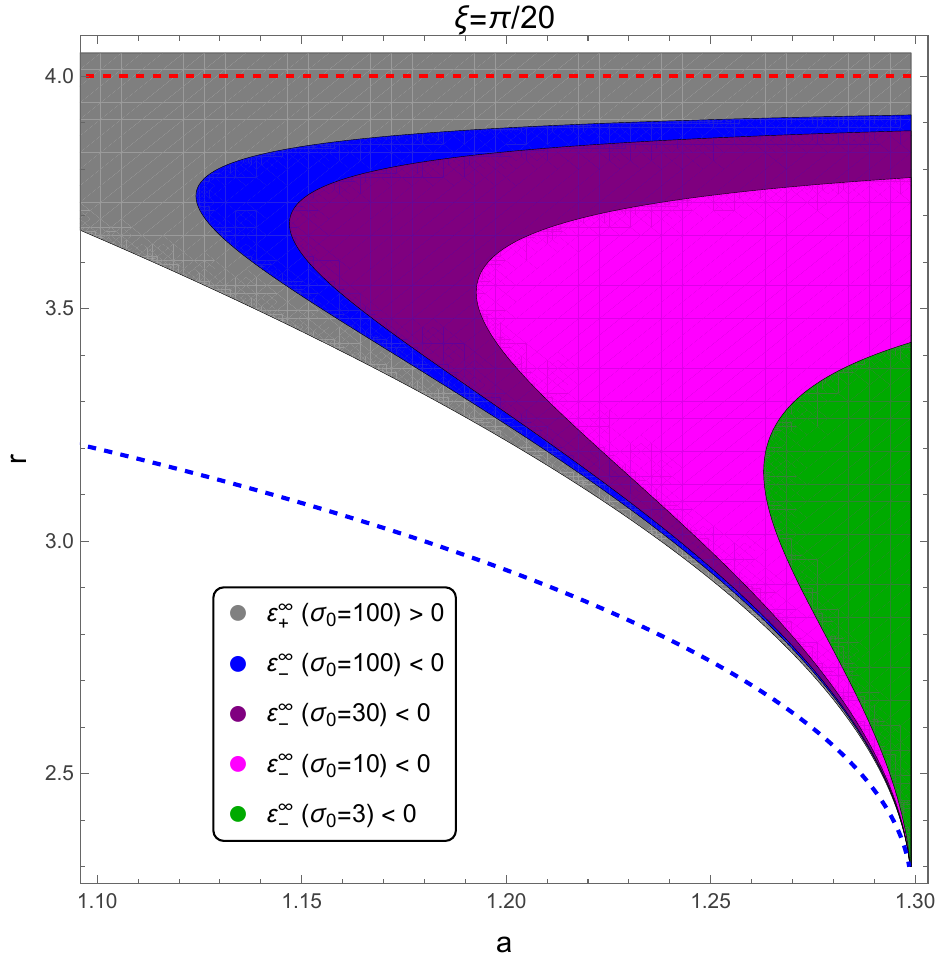}
  &
  \includegraphics[width=0.48\textwidth]
  {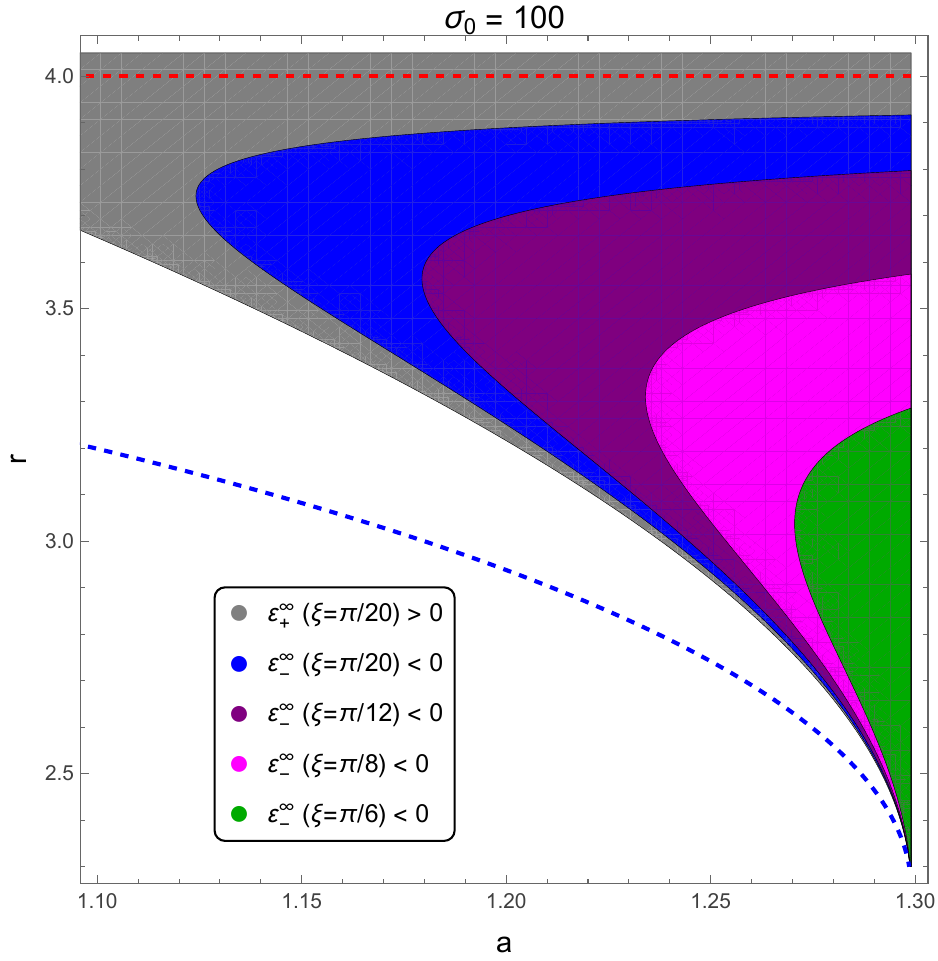}
  \\[2mm]
  \includegraphics[width=0.48\textwidth]
  {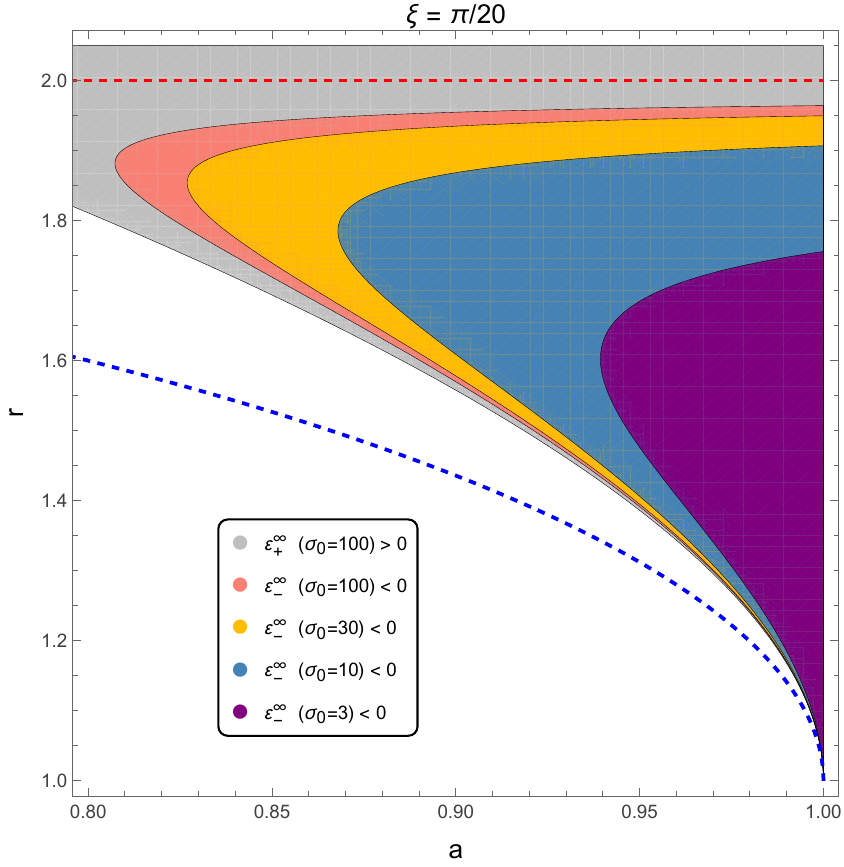}
  &
  \includegraphics[width=0.48\textwidth]
  {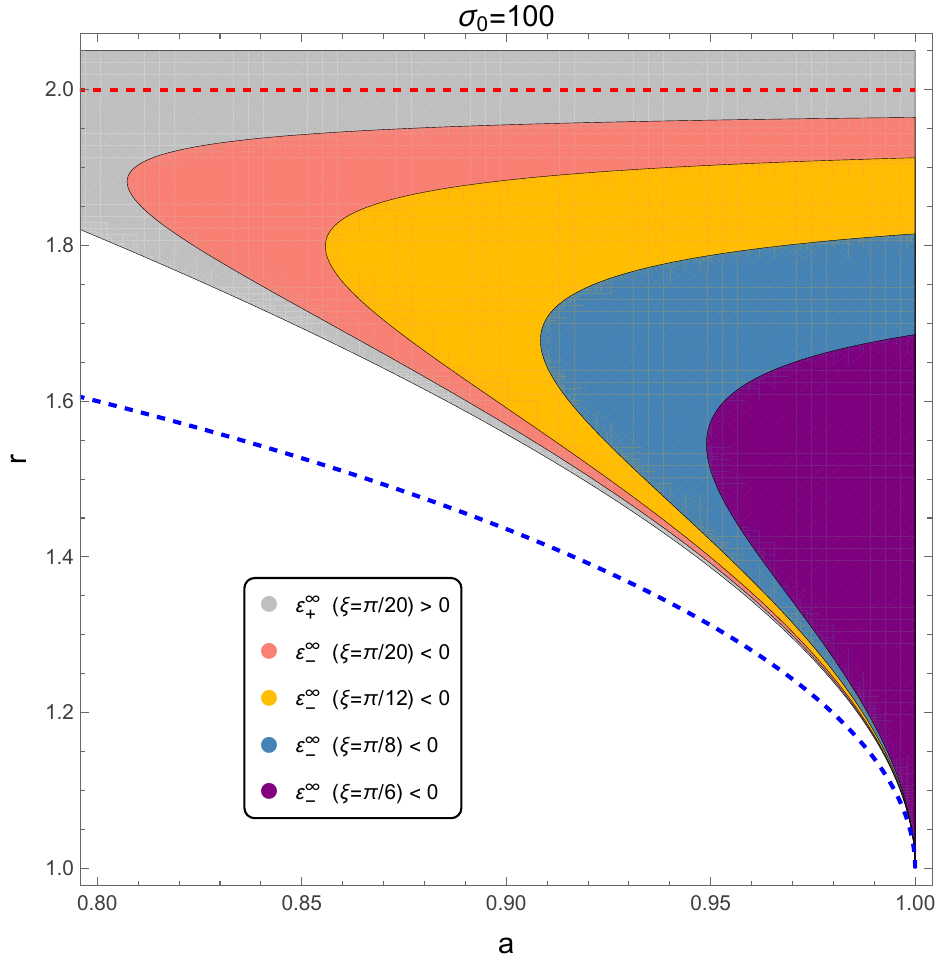}
\end{tabular}

\caption{
For rotating $D=6$ (top-row) and $D=7$ (bottow-row) dimensional pure Lovelock BHs with a single rotation, ($a$--$r$) parameter-space regions in which energy extraction is allowed based on the energy-at-infinity conditions $\epsilon^{\infty}_{+}>0$  and $\epsilon^{\infty}_{-}<0$ provided for different values of magnetization parameteres ($\sigma_0=3-100$) when the orientation angle value is fixed ($\xi=\pi/20$) and different values of the orientation angle ($\xi=\pi/20-\pi/6$) when the magnetization factor is fixed ($\sigma_0=100$). The blue dashed lines show the outer event horizons and the red dashed lines mark the outer boundaries of the corresponding ergosphere of BHs.
}
\label{fig:6D7Dregions}
\end{figure*}
\begin{figure*}[ht]
\centering
\begin{tabular}{cc}
  \includegraphics[width=0.48\textwidth]
  {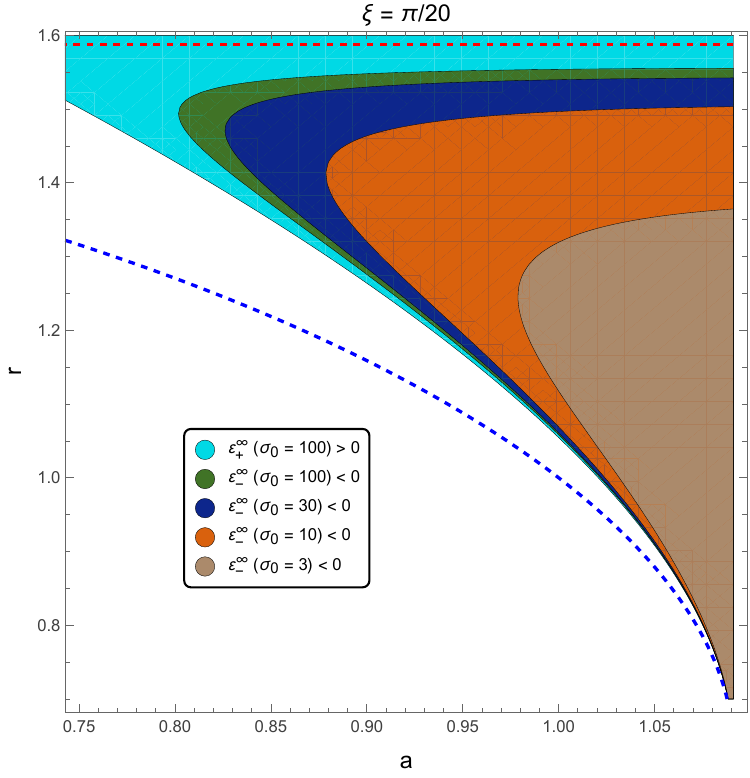}
  &
  \includegraphics[width=0.48\textwidth]
  {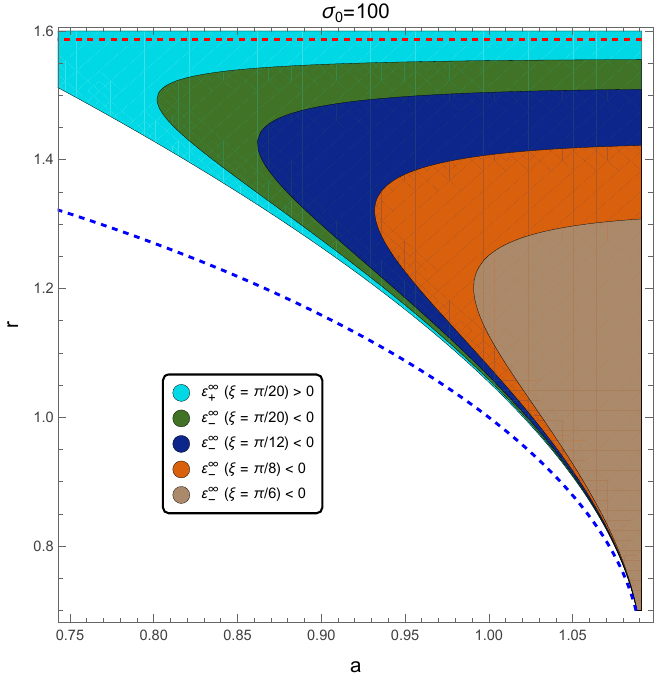}
  \\[2mm]
  \includegraphics[width=0.48\textwidth]
  {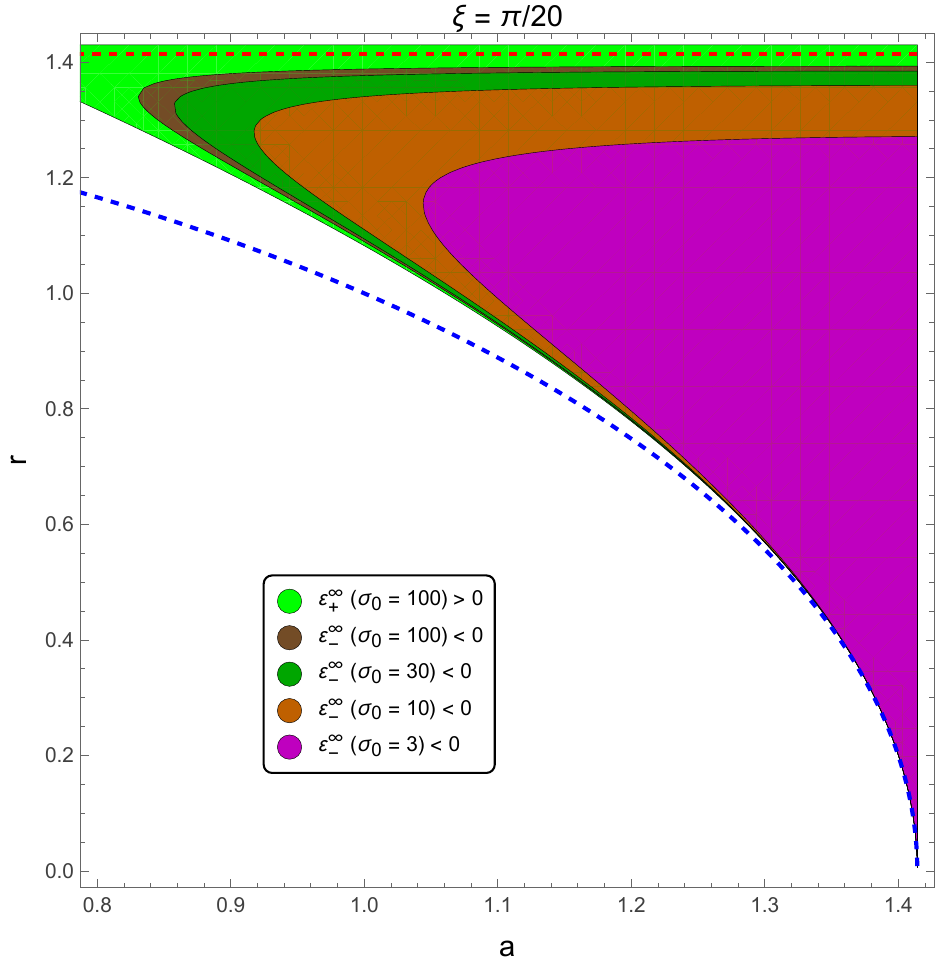}
  &
  \includegraphics[width=0.48\textwidth]
  {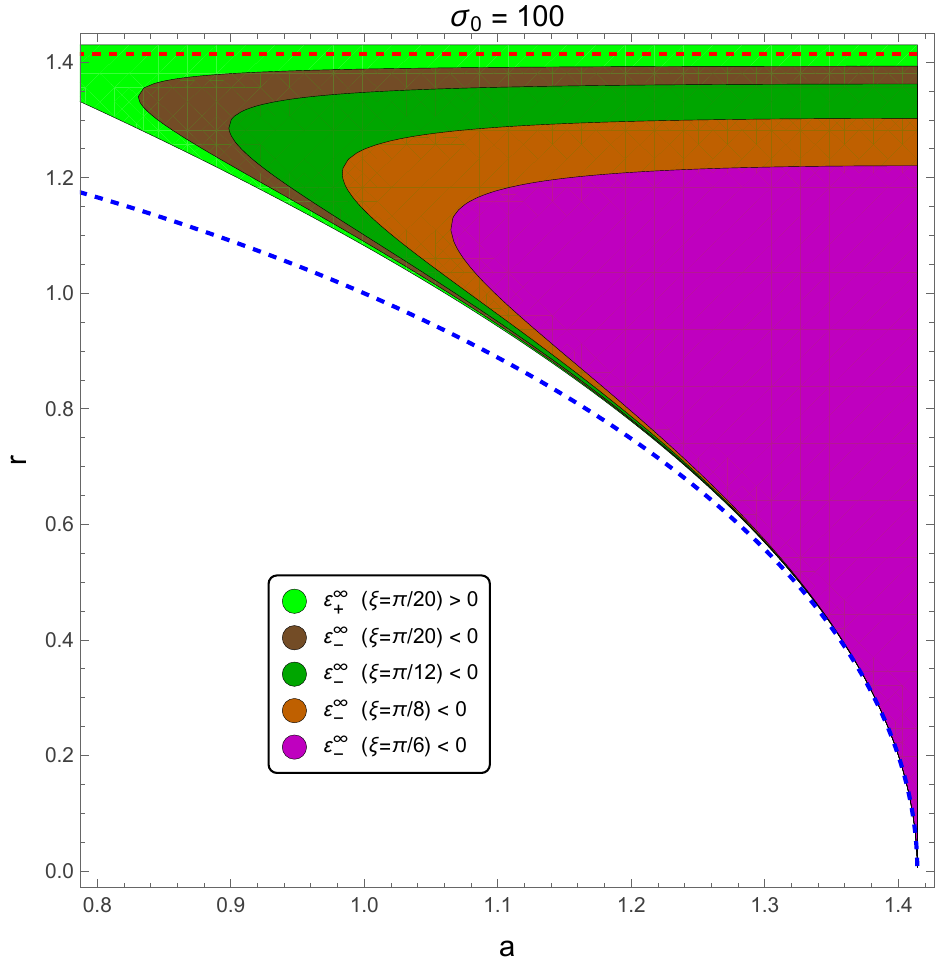}
\end{tabular}

\caption{
For $D=8$ (top-row) and $D=9$ (bottow-row) dimensional pure Lovelock BHs with a single rotation, ($a$--$r$) parameter-space regions in which energy extraction is allowed based on the energy-at-infinity conditions $\epsilon^{\infty}_{+}>0$ and $\epsilon^{\infty}_{-}<0$ provided for different values of magnetization parameteres ($\sigma_0=3-100$) when the orientation angle value is fixed ($\xi=\pi/20$) and different values of the orientation angle ($\xi=\pi/20-\pi/6$) when the magnetization factor is fixed ($\sigma_0=100$). The blue dashed lines show the outer event horizons and the red dashed lines mark the outer boundaries of the corresponding ergospheres of the BHs.
}
\label{fig:8D9Dregions}
\end{figure*}
  \begin{eqnarray}
      ds^2=-d\hat{t}^2+\sum_{i=1}^k(d\hat{x^i})^2=\eta_{\mu \nu}\hat{dx^{\mu}}\hat{dx^{\nu}}\, . 
  \end{eqnarray}
The transformation relating ZAMO frame ($\hat{t}$, $\hat{x^i}$) and the Boyer-Lindquist coordinates ($t$, $x^i$) is given as
\begin{eqnarray}
d\hat{t}=\alpha dt \mbox{~~and~~} d\hat{x^i}=\sqrt{g_{ii}}dx^i-\alpha \beta^{i}dt\, ,
\end{eqnarray}
where $\alpha$ denotes the lapse function, which relates the coordinate time to the proper time measured by a ZAMO: $\alpha=\sqrt{-g_{tt}+\frac{g_{\phi t}^2}{g_{\phi \phi}}}$ and $\beta^{i}$ refers to the shift vector, which accounts for the frame-dragging effect in rotating spacetime: $\beta^{\phi}=\frac{\sqrt{g_{\phi\phi}}\,\omega^{\phi}}{\alpha}$. Because we consider the single rotation configuration case of higher-dimensional BHs, only one independent rotation parameter is present, and a single nonvanishing off-diagonal metric component, $g_{t\phi}$, remains. Considering that the shift vector is expressed by the time-space components of the metric and only $g_{t\phi}$ is nonzero, the shift vector's just azimuthal component exists. Furthermore, we analyse the process in the equatorial plane and it keeps the remaining angular coordinates, so that the shift vectors' form becomes as $\beta^{i}=\left(0,\ldots,0,\beta^{\phi}\right).$ $\omega^{\phi}=-{g_{\phi t }}/{g_{\phi \phi}}$ is considered angular velocity of frame dragging.

Further we assess the viability of magnetic reconnection mechanism by examining accelerated and decelerated plasma energies in the ergoregions of the BHs. The energy-momentum tensor, under one fluid approximation can be written as
\begin{eqnarray}
    T^{\mu \nu}=pg^{\mu \nu}+{\mathit{w}}U^{\mu}U^{\nu}+F^{\mu}_{\delta}F^{\nu \delta}-\frac{1}{4}g^{\mu \nu}F^{\rho \delta}F_{\rho \delta}\, , 
\end{eqnarray}
where $p$ and $\omega$ are considered plasma pressure and enthalpy density while $U^{\mu}$ and $F^{\mu\nu}$ refer to the four-velocity and the electromagnetic field tensor correspondingly. The enthalpy density can be defined as \cite{Koide08ApJ}
\begin{equation}
\begin{aligned}
\mathit{w} = e_{\mathrm{int}} + p,
\qquad
e_{\mathrm{int}} = \frac{p}{\Gamma-1}+\rho c^2 ,
\end{aligned}
\end{equation}
where $e_{\mathrm{int}}$ is considered plasma energy density in its local rest frame, the first term in the expression of $e_{\mathrm{int}}$ is thermal internal energy density and the second term defines rest-mass energy density, furthermore,  $\Gamma$ and $\rho$ refer to the adiabatic index and proper mass density of plasma respectively. To derive the energy of the plasma as measured by a static observer at infinity, we express the "energy-at-infinity" quantity as
\begin{equation}
    e^\infty = -\alpha\, g_{\mu 0}\, T^{\mu 0}=\alpha \hat{e}+\alpha \beta^{\phi}\hat{P}^{\phi}\, ,
\end{equation}
here $\hat{e}$ and $\hat{P}^{\phi}$ refer to the total energy density and azimuthal component of momentum density correspondingly and read as
\begin{eqnarray}
    \hat{e}={\mathit{w}}\hat{\gamma}^2-p+\frac{1}{2}(\hat{B}^2+\hat{E}^2) ,\\\
   \hat{P}^{\phi}={\mathit{w}}\hat{\gamma}^2\hat{v}^{\phi}+(\hat{B} \times \hat{E})^{\phi}\, ,
\end{eqnarray}
here $\hat{v}^{\phi}$ refers to the $\phi$-component of plasma outflow velocity in the ZAMO, $\hat{\gamma}=\hat{U}^0=({1-\sum_{i=1}^3(\hat{v}^i)^2})^{-1/2}$ is the relativistic Lorentz factor and electric/magnetic field components are $\hat{E}^i=\hat{F}^{i0}$ and $\hat{B}^i=\epsilon^{ijk}\hat{F}_{jk}/2$. 
The energy density measured at infinity consists of hydrodynamic and electromagnetic contributions:
\begin{align}
e^{\infty}
&= e_{\mathrm{hyd}}^{\infty}
 + e_{\mathrm{em}}^{\infty},
\\
e_{\mathrm{hyd}}^{\infty}
&= \alpha \hat{e}_{\mathrm{hyd}}
 + \alpha \beta^{\phi} w \hat{\gamma}^{2}
   \hat{v}^{\phi},
\\
e_{\mathrm{em}}^{\infty}
&= \alpha \hat{e}_{\mathrm{em}}
 + \alpha \beta^{\phi}
   \left(\hat{\mathbf{B}}\times\hat{\mathbf{E}}\right)_{\phi},
\end{align}
in the above equations, $\hat{e}_{\mathrm{em}}$ and $\hat{e}_{\mathrm{hyd}}$ are considered the electromagnetic and hydrodynamic energy densities in the ZAMO frame respectively and read as
\begin{equation}
\hat{e}_{\rm em}
= \frac{\hat{B}^{2}+\hat{E}^{2}}{2}
\qquad \text{and} \qquad
\hat{e}_{\rm hyd}
= \mathit{w}\hat{\gamma}^{2}-p .
\end{equation}
We should note that, using the approximations of Comisso and Asenjo in~\cite{Comisso21}, the plasma in our examination is taken as non-compressible and adiabatic, hence, the electromagnetic term of the energy density is negligible in contrast to the hydrodynamic term. Therefore we can define the total energy density as
\begin{equation}
    e^{\infty}=e^{\infty}_{hyd}=\alpha [{\mathit{w}}\hat{\gamma}(1+\beta^{\phi}\hat{v}^{\phi})-\frac{ p}{\hat{\gamma}}]\, .
\end{equation}
If we are given any vector quantity $f$, the coordinate transformations of the vector from Boyer-Lindquist to the ZAMO can be written as
\begin{align}
\hat{f}_0
&=
\frac{f_0}{\alpha}
+
\sum_{i=1}^{3}\frac{\beta^i}{g_{ii}}\,f_i
\qquad \text{and} \qquad
\hat{f}_i
=
\frac{f_i}{\sqrt{g_{ii}}};
\\[-0.3em]
\hat{f}^{0}
&=
\alpha f^{0}
\qquad \text{and} \qquad
\hat{f}^{i}
=
\sqrt{g_{ii}}\,f^{i}
-
\alpha\beta^{i}f^{0}.
\end{align}
Using the above transformations, Keplerian velocity of the co-rotating plasma read as
\begin{eqnarray}
 \hat{v}_K&=&\frac{d\hat{x}^{\phi}}{d\hat{t}}
 =\frac{\sqrt{g_{\phi \phi}}}{\alpha}\Omega_K-\beta^{\phi}\, , 
 \end{eqnarray}
 where $\Omega_K$ is Keplerian angular velocity of the plasma and expressed as
\begin{eqnarray}
\Omega_K=\frac{d\phi}{dt}=\frac{-\partial_r g_{t\phi}\pm\sqrt{(\partial_r g_{t\phi})^2-(\partial_r g_{tt})(\partial_r g_{\phi\phi})}}{\partial_r g_{\phi\phi}}\, .
\end{eqnarray}
According to the Comisso-Asenjo consideration~\cite{Comisso21}, the reconnecting layer is situated within the azimuthal direction and the current density lies radially. Depending on the upstream plasma magnetization parameter $\sigma_0$, the plasma outflow velocity and its Lorentz gamma factor read as
\begin{equation}
v_{out}\approx (\frac{\sigma_0}{1+\sigma_0})^{1/2} ,
\end{equation}
\begin{equation}
\gamma_{out}= (1-v_{out}^2)^{-1/2}\approx (1+\sigma_0)^{1/2}\, ,
\end{equation}
where the magnetization parameter is defined as $\sigma_0=B_0^2/\omega_0$, and together with $B_0$, it characterizes the large-scale magnetic field, $\omega_0$ is the enthalpy density of plasma as mentioned above.
The azimuthal components of corotating (+) and counter-rotating (-) outflow velocities of plasma are written as 
\begin{equation}
v_{\pm}^{\phi}=\frac{\hat{v}_K \pm v_{out}\cos{\xi}}{1 \pm \hat{v}_K v_{out}\cos{\xi}}\, ,
\end{equation}
where $\xi$ refers to the angle between the outflow velocity components and represents the orientation angle of the reconnecting magnetic field. Combining the above equations, we obtain the final form of energy at infinity per enthalpy $e^{\infty}/\omega$~\cite{Comisso21}:
\begin{eqnarray}\label{Eq:essential}
 \epsilon^{\infty}_{\pm}&=&\alpha \hat{\gamma}_K \Bigg[(1+\beta^{\phi}\hat{v}_K)\sqrt{1+\sigma_0}\pm \cos{\xi}(\beta^{\phi}+\hat{v}_K)\sqrt{\sigma_0}\nonumber\\
 &-&\frac{\sqrt{1+\sigma_0}\mp \cos{\xi}\hat{v}_K\sqrt{\sigma_0}}{4\hat{\gamma}^2(1+\sigma_0-\cos^2{\xi}\hat{v}_K^2\sigma_0)}\Bigg]\, ,  
\end{eqnarray}
here, $\epsilon^{\infty}_{-}$ and $\epsilon^{\infty}_{+}$ denote the energies per unit enthalpy of the decelerated and accelerated plasma, respectively, as measured by an observer at infinity. For energy extraction from a BH within its ergosphere, the energies at infinity of the decelerated and accelerated plasma components must satisfy the following conditions:
\begin{eqnarray}
 \epsilon_{-}^{\infty}<0\, , 
 \end{eqnarray}
 \begin{eqnarray}
 \Delta \epsilon_{+}^{\infty}=\epsilon_{+}-\left(1-\frac{\Gamma}{\Gamma-1}\frac{p}{\mathit{w}}\right)=\epsilon_{+}^{\infty}>0\, ,
\end{eqnarray}
where $\Gamma$ defines the state of relativistically hot plasma and equals (4/3).
\begin{figure*}[ht]
\centering
\begin{tabular}{cc}
  \includegraphics[width=0.48\textwidth]
  {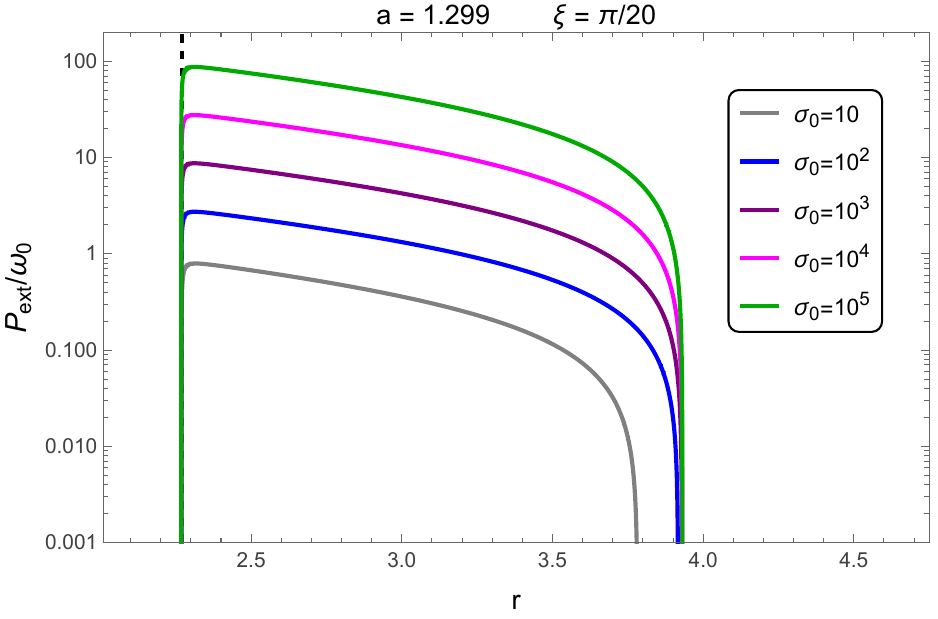}
  &
  \includegraphics[width=0.48\textwidth]
  {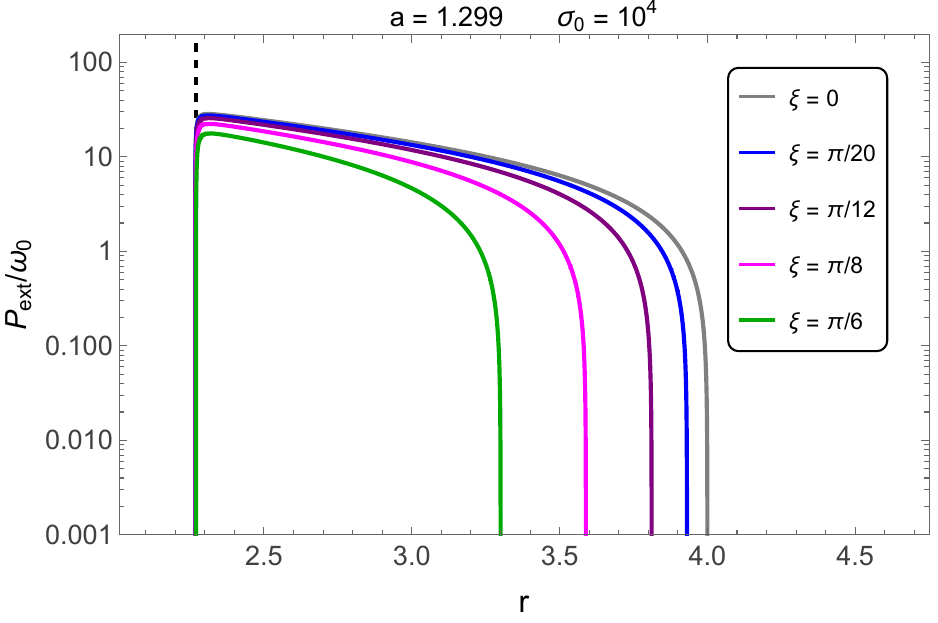}
  \\[2mm]
  \includegraphics[width=0.48\textwidth]
  {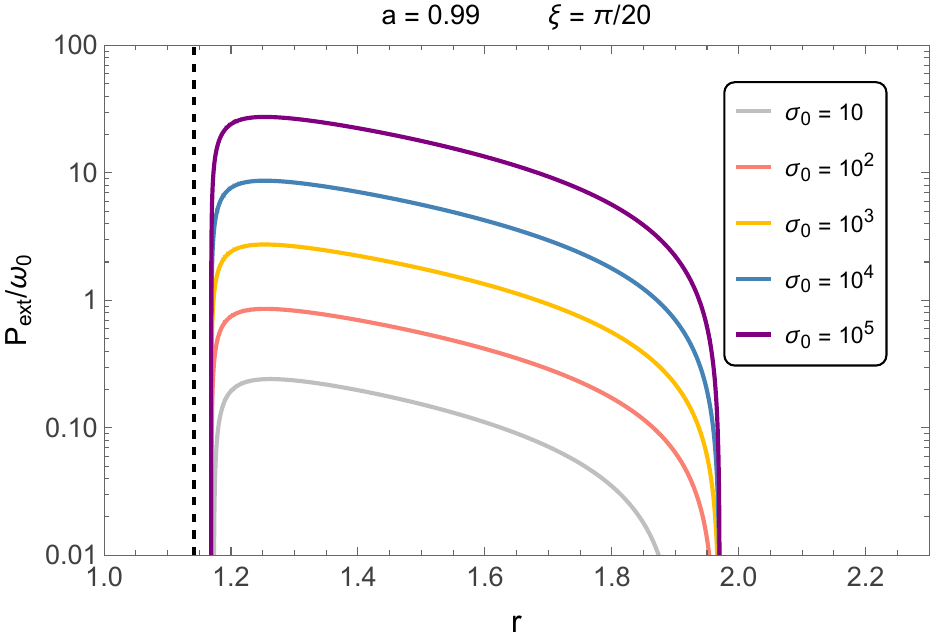}
  &
  \includegraphics[width=0.48\textwidth]
  {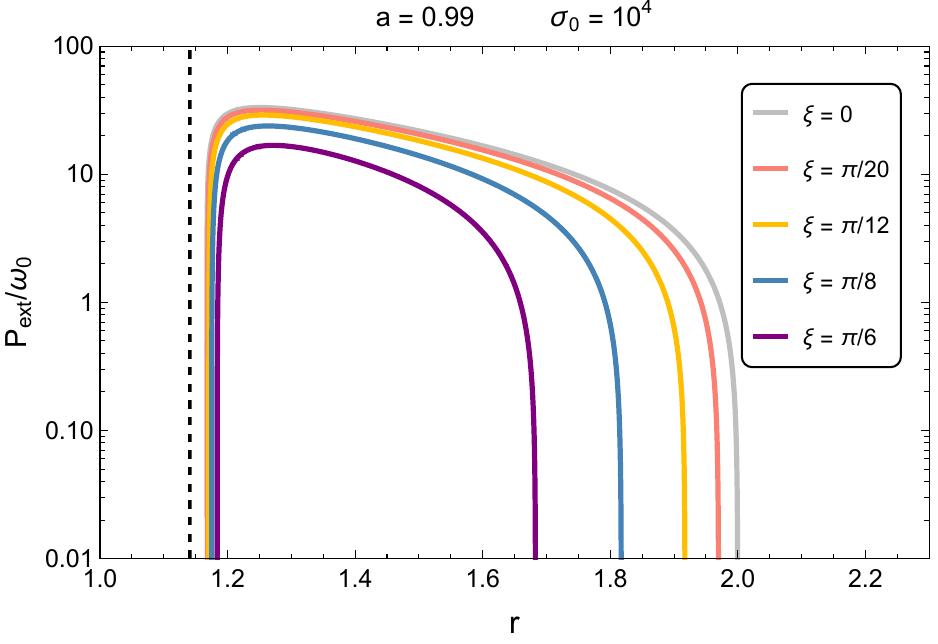}
\end{tabular}

\caption{For maximally rotating $D=6$ (top-row) and $D=7$ (bottom-row) dimensional BHs with a single rotation, the extracted power per unit enthalpy density as a function of reconnection occuring location provided for different magnetization parameters $\sigma_0\in\{10,10^2,10^3,10^4,10^5\}$ when the orientation angle is $\xi=\pi/20$ and for different values of the orientation angle $\xi\in\{0,\pi/20,\pi/12,\pi/8,\pi/6\}$ when the magnetization is fixed to $\sigma_0=10^4$
}
\label{fig:6D7DPower}
\end{figure*}
\begin{figure*}[ht]
\centering
\begin{tabular}{cc}
  \includegraphics[width=0.48\textwidth]
  {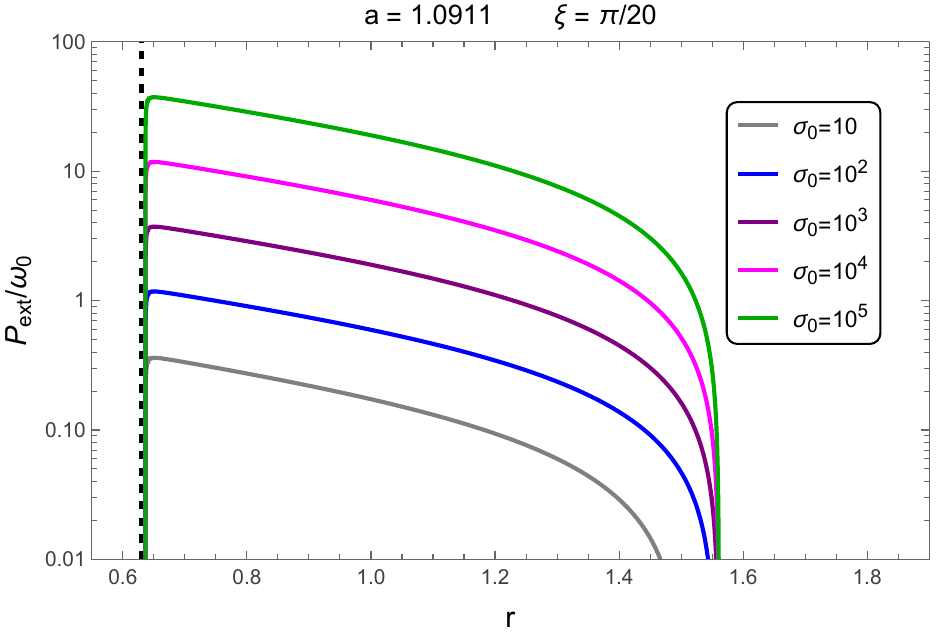}
  &
  \includegraphics[width=0.48\textwidth]
  {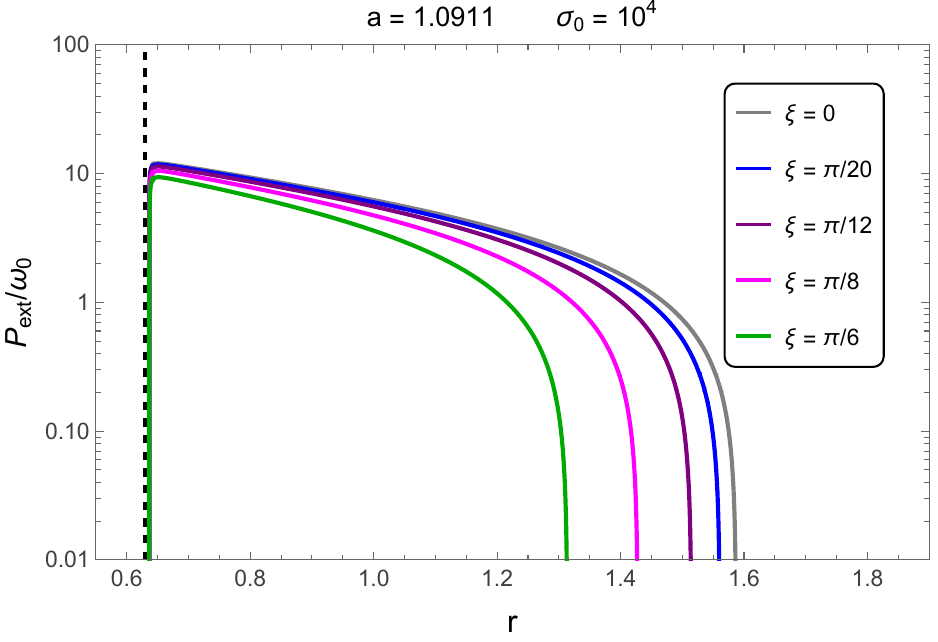}
  \\[2mm]
  \includegraphics[width=0.48\textwidth]
  {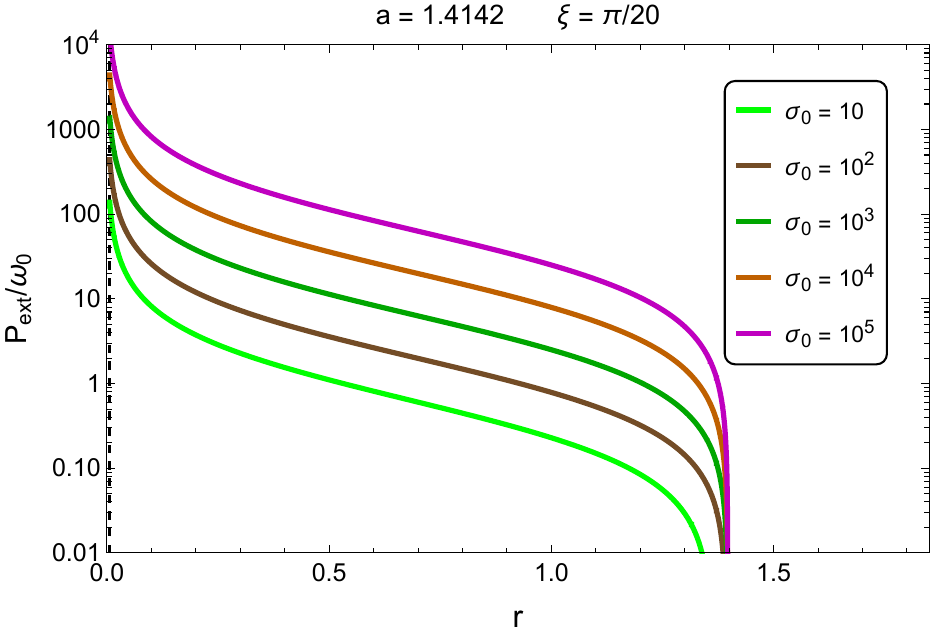}
  &
  \includegraphics[width=0.48\textwidth]
  {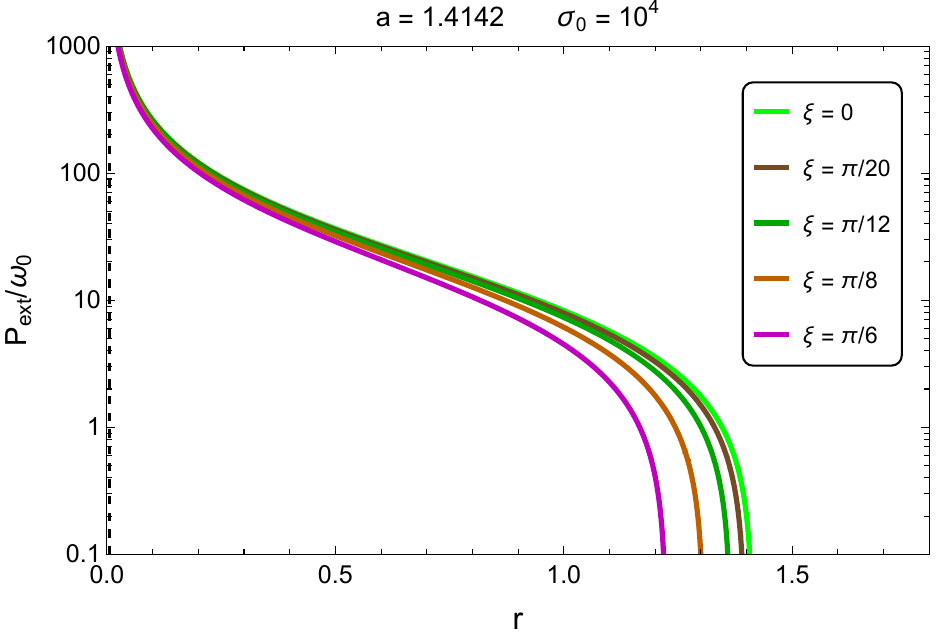}
\end{tabular}

\caption{For maximally rotating $D=8$ (top-row) and $D=9$ (bottom-row) dimensional BHs with a single rotation, the extracted power per unit enthalpy density as a function of reconnection occurring location provided for different magnetization parameters $\sigma_0\in\{10,10^2,10^3,10^4,10^5\}$ when the orientation angle is $\xi=\pi/20$ and for different values of the orientation angle $\xi\in\{0,\pi/20,\pi/12,\pi/8,\pi/6\}$ when the magnetization is fixed to $\sigma_0=10^4$
}
\label{fig:8D9DPower}
\end{figure*}
We have applied the Comisso–Asenjo mechanism to pure Lovelock BHs in $D=6,\,7,\,8,\,9$ dimensions and analyzed the energies at infinity of the accelerated and decelerated plasma components in each case using Eq.~(\ref{Eq:essential}) and its dependence on the key reconnection parameters. Figure~\ref{Fig:energyatinf} illustrates the relationship between the magnetization parameter and the energies at infinity per enthalpy of the accelerated and decelerated plasma components for pure Lovelock BHs in $D=6,\,7,\,8,\,9$ dimensions, with the spin parameters approaching their maximum values, $\xi \to 0$, and $r$ approaching the corresponding outer-horizon radius. It can be seen from the figure that, in all cases, the energies at infinity of both accelerated and decelerated plasma components depend strongly on the magnetization parameter $\sigma_0$, with higher magnetization values being required for efficient energy extraction, hence, $\sigma\gg1$ should be satisfied.

Figures~\ref{fig:6D7Dregions} and~\ref{fig:8D9Dregions} illustrate the ($a$--$r$) parameter-space regions for pure Lovelock BHs in $D=6,\,7,\,8,\,9$ dimensions, respectively. Here, $a$ denotes the dimensionless spin parameter, while $r$ is the radial location of magnetic reconnection. These regions represent the viability of energy extraction through the Comisso--Asenjo mechanism based on the conditions $\epsilon^{\infty}_{+}>0$ and $\epsilon^{\infty}_{-}<0$ for the accelerated and decelerated plasma components, respectively. Right panels of the figures are made for reconnecting magnetic field orientation angle $\xi=\pi/20$ and varying quantities of magnetization parameter $\sigma_0\in\{3,10,30,100\}$. Left panels of the figures are done for the magnetization factor $\sigma_0=100$ and different values of the orientation angle $\xi\in\{\pi/6,\pi/8,\pi/12,\pi/20\}$. The blue and red dashed lines represent the outer-horizon limits and the outer boundaries of the ergoregions of the BHs, respectively. We can see from Figs.~\ref{fig:6D7Dregions} and \ref{fig:8D9Dregions} that, for each dimension, the rotation parameter and the radial location of magnetic reconnection vary over different ranges, which are strongly constrained by the corresponding horizon and ergoregion boundaries of the BHs. However, the dependence of the parameter-space regions on the magnetization parameter and the orientation angle shows very similar behavior for the BHs in all $D=6,\,7,\,8,\,9$ dimensions. Therefore, we can analyze these dependencies collectively. As shown in the left panels of both Figs.~\ref{fig:6D7Dregions} and \ref{fig:8D9Dregions}, increasing the magnetization parameter $\sigma_0\in\{3,10,30,100\}$ leads to expanding the parameter-space regions in which BH energy can be extracted through the Comisso--Asenjo mechanism toward larger reconnection radii $r$ and lower values of the dimensionless rotation parameter $a$. Increasing the plasma magnetization factor allows the energy-extraction regions to expand toward the outer boundaries of the ergoregions. 

As shown in the right panels of both Figs.~\ref{fig:6D7Dregions} and \ref{fig:8D9Dregions}, the parameter-space regions in which energy extraction via the Comisso--Asenjo mechanism is viable broaden toward larger values of the reconnection location $r$ and lower values of the dimensionless spin $a$ as the reconnection orientation angle $\xi\in\{\pi/6,\pi/8,\pi/12,\pi/20\}$ decreases. Only the azimuthal component of the plasma outflow velocity has an important role to the extraction of rotational energy; therefore, the energy-extraction regions shrink when the orientation angle increases. As the orientation angle value approaches zero, the radial location of magnetic reconnection, $r$, approaches the outer boundary of the ergosphere and it leads to more effective rotational energy extraction. In Figs.~\ref{fig:6D7Dregions} and \ref{fig:8D9Dregions}, blue and red dashed lines represent the limits of outer event horizons and the outer ergoregion boundaries of the BHs respectively. From a physical perspective, BH rotational energy extraction is possible within these limits. Since all of the viable parameter-space regions obtained in our analysis lie within them, the conditions considered here permit energy extraction from pure Lovelock BHs in all dimensions $D=6,\,7,\,8,\,9$.

It should be noted that, for the $D=7$ dimensional pure Lovelock BH, both the dependence of the accelerated/decelerated plasma energies at infinity per unit enthalpy on the magnetization factor (see Fig.~\ref{Fig:energyatinf}) and the corresponding parameter-space regions (see Fig.~\ref{fig:6D7Dregions}) are closely similar to those obtained for energy extraction from a four-dimensional Kerr BH in Einstein gravity through magnetic reconnection~\cite{Comisso21}. This similarity arises from the similar geometric structures of the event horizon and ergosphere in both cases. 

In the following section, we wish to analyze the power extracted from pure Lovelock BHs in  $D=6,\,7,\,8,\,9$ dimensions and the efficiency of Comisso-Asenjo MR mechanism. 
\begin{figure*}[ht]
\centering
\begin{tabular}{cc}
  \includegraphics[width=0.48\textwidth]
  {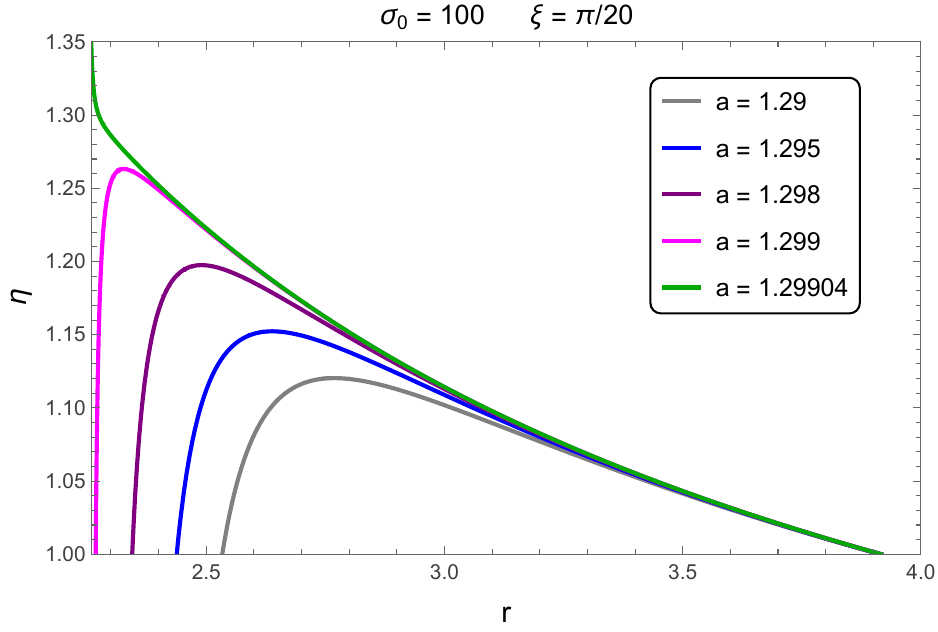}
  &
  \includegraphics[width=0.48\textwidth]
  {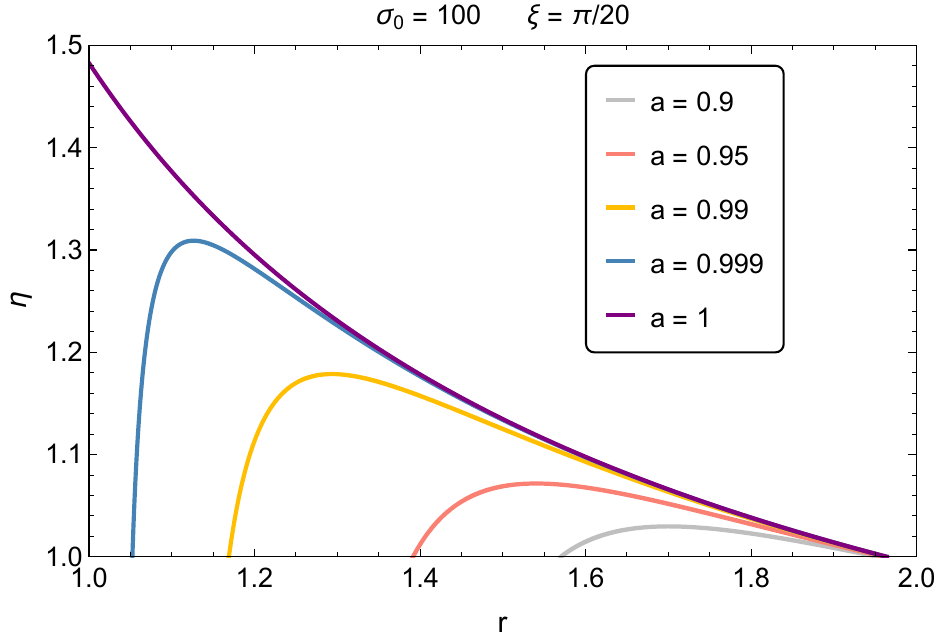}
  \\[2mm]
  \includegraphics[width=0.48\textwidth]
  {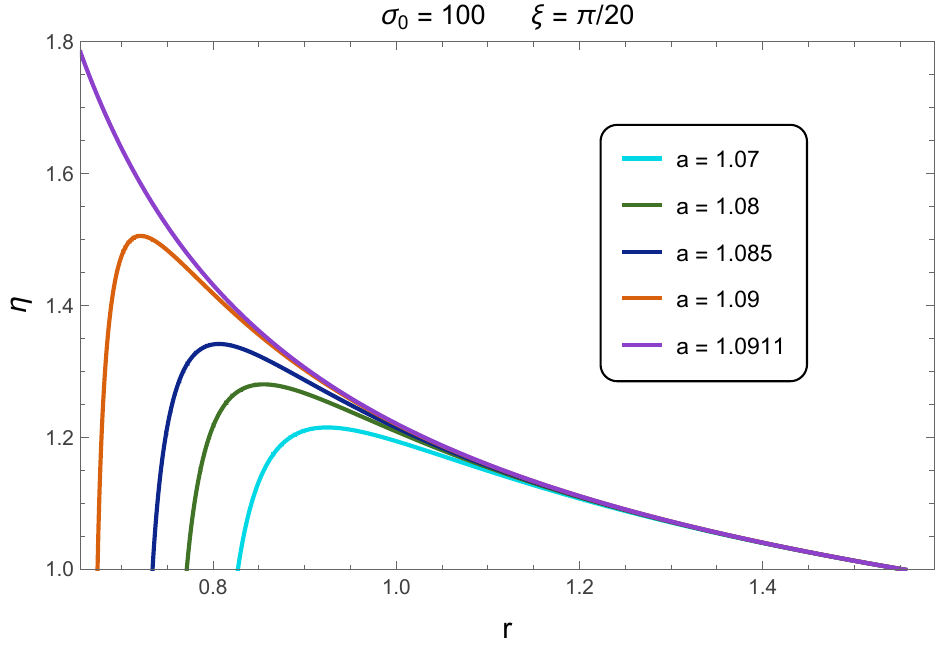}
  &
  \includegraphics[width=0.48\textwidth]
  {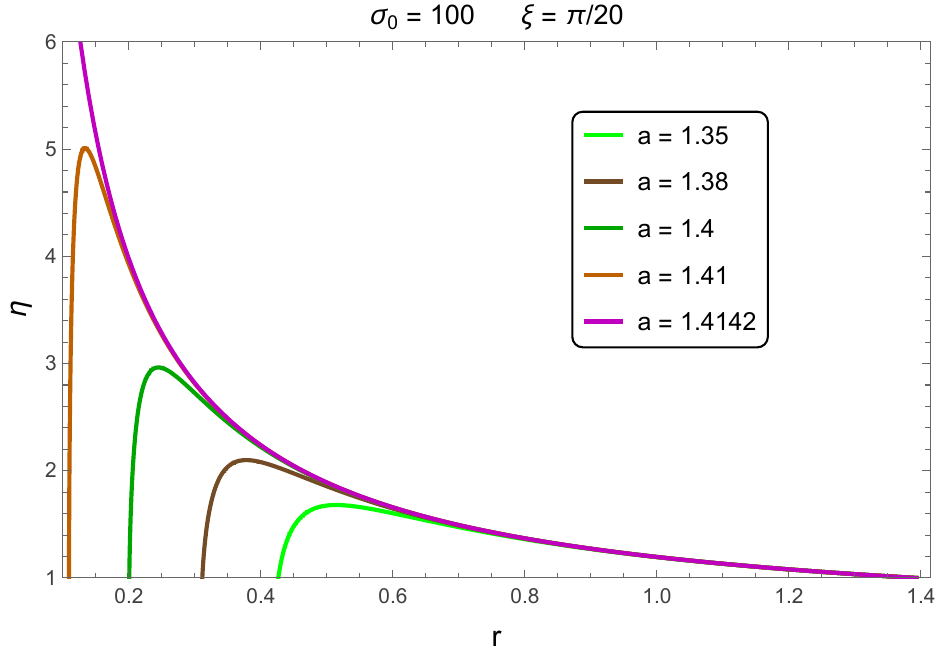}
\end{tabular}

\caption{The energy extraction efficiency via the Comisso--Asenjo mechanism, $\eta$, is shown as a function of the radial location of magnetic reconnection, $r$, for $D=6$ (upper-left panel), $D=7$ (upper-right panel), $D=8$ (lower-left panel), and $D=9$ (lower-right panel) dimensional rotating pure Lovelock BHs with a single rotation. The magnetization parameter and the orientation angle are fixed at $\sigma_0=100$ and $\xi=\pi/20$, respectively, while the spin parameter is chosen to be near its maximum value in each dimension.}
\label{fig:efficiency}
\end{figure*}
\begin{figure*}[ht]
\centering
\begin{tabular}{cc}
  \includegraphics[width=0.48\textwidth]
  {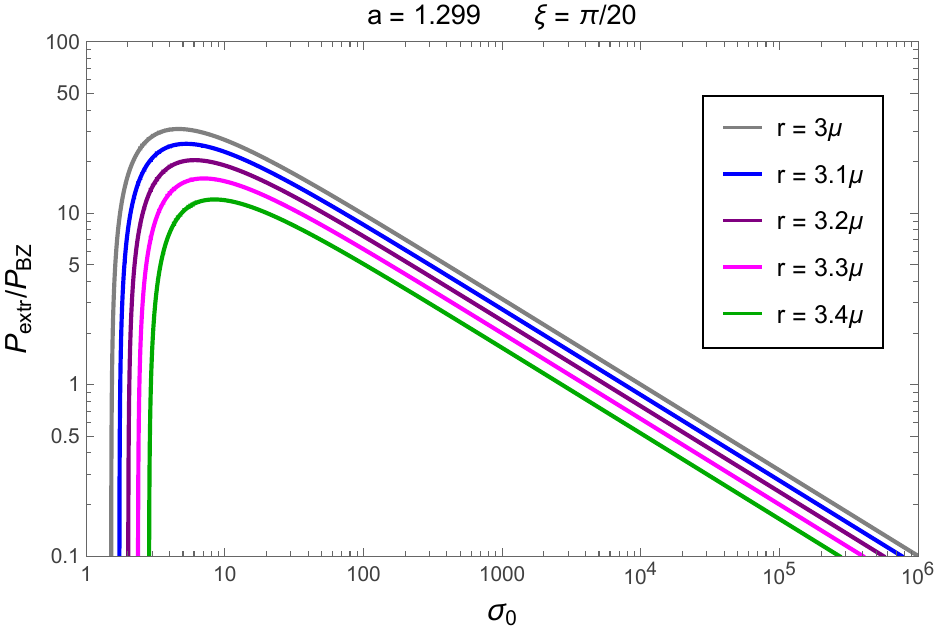}
  &
  \includegraphics[width=0.48\textwidth]
  {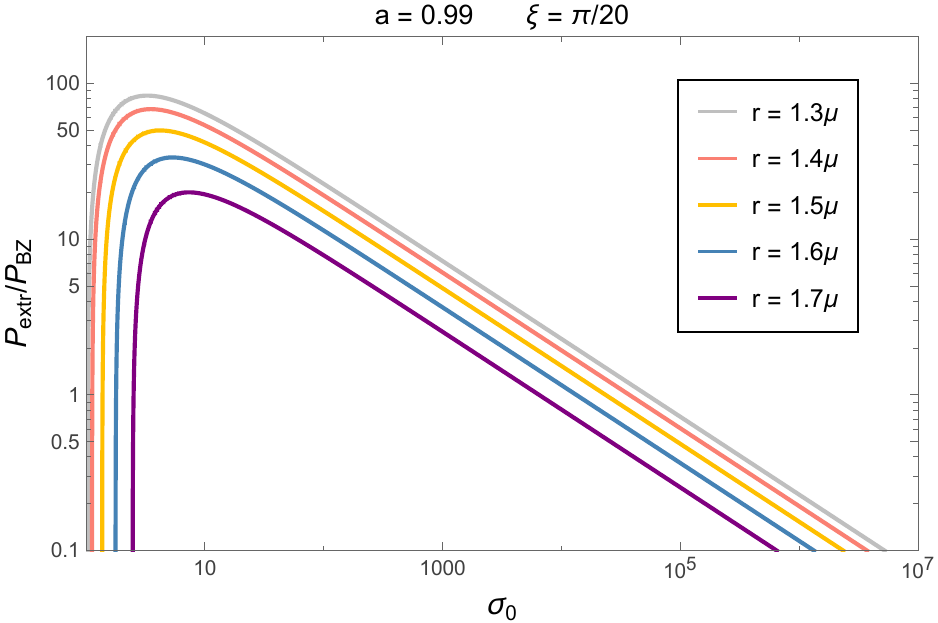}
  \\[2mm]
  \includegraphics[width=0.48\textwidth]
  {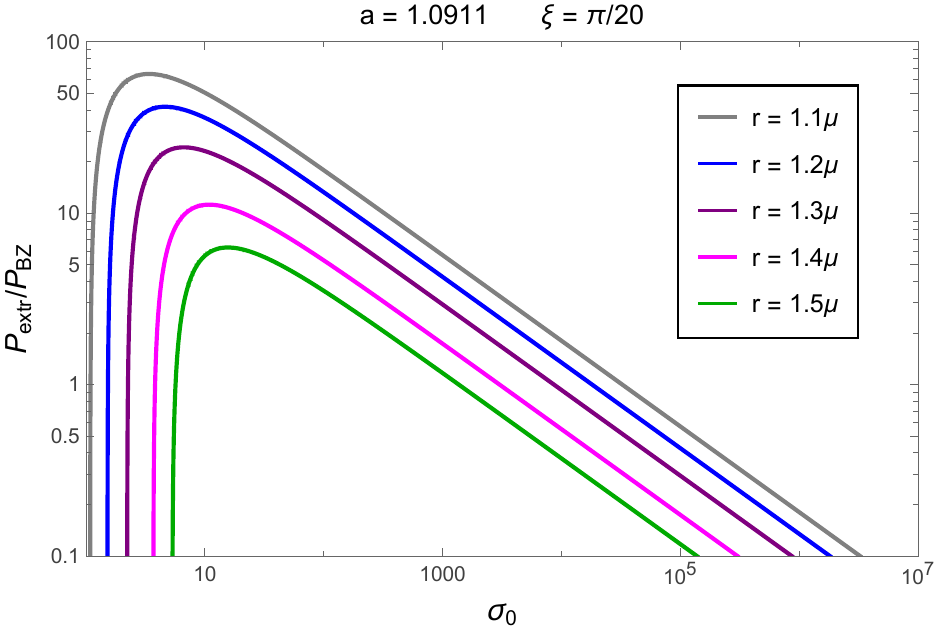}
  &
  \includegraphics[width=0.48\textwidth]
  {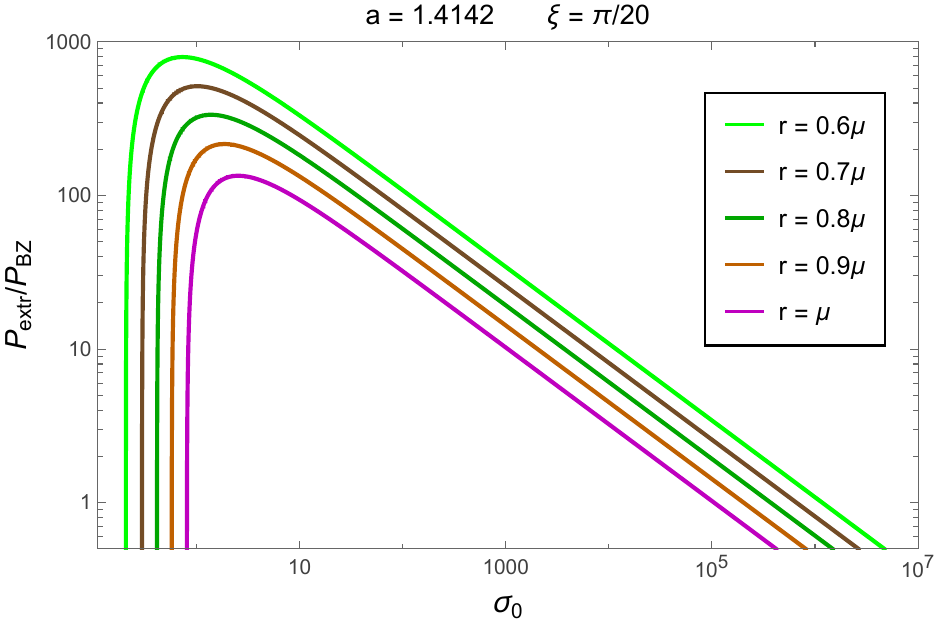}
\end{tabular}

\caption{Comparison of the power extracted via the Comisso--Asenjo and Blandford--Znajek mechanisms, expressed as $P_{\mathrm{ext}}/P_{\mathrm{BZ}}$, as a function of the magnetization parameter. The orientation angle is fixed at $\xi=\pi/20$ for maximally rotating pure Lovelock BHs with a single rotation in $D=6$ (upper-left panel), $D=7$ (upper-right panel), $D=8$ (lower-left panel), and $D=9$ (lower-right panel) dimensions. The curves are made for different radial locations of magnetic reconnection.
}
\label{fig:PowerMRBZ}
\end{figure*}

\section{\label{Sec:MP-Power-MR} THE EXTRACTED POWER AND EFFICIENCY OF THE COMISSO-ASENJO MR MECHANISM}

Since energy extraction is possible under the conditions discussed above, we can investigate the extraction of rotational energy from the higher-dimensional pure Lovelock BHs via the Comisso-Asenjo MR mechanism \cite{Comisso21}. As stated earlier, in magnetic reconnection, the decelerated plasma with negative energy falls into the BH and the accelerated one takes rotational energy of the BH and goes to infinity. The power taken out by the escaping plasma from the BH via the Comisso-Asenjo MR mechanism can be written as
\begin{eqnarray}
    P_{ext}=-\epsilon_{-}^{\infty}{\mathit{w}}_0 A_{in} U_{in}\, ,
\end{eqnarray}
as already known, $\epsilon_{-}^{\infty}$ is the energy at infinity per unit enthalpy of the decelerated plasma, $\omega_0$ is considered the density of enthalpy, $A_{in}$ is the cross-sectional area and $U_{in}$ is the regime-dependant coefficient of the magnetic reconnection. We consider the mechanism occurs in collisionless regime, therefore $U_{in}={\cal O}(10^{-1})$(for collisional regimes $U_{in}$ is taken as ${\cal O}(10^{-2})$). For maximally rotating BHs cross-sectional parameter can be written as $A_{in}$= $(r_{st}^2-r_H^2)$. 

Figures~\ref{fig:6D7DPower} and~\ref{fig:8D9DPower} represent the rotational power extracted from the BHs through the Comisso--Asenjo mechanism per unit enthalpy density as a function of the radial location of magnetic reconnection for $D=6$ and $D=7$, $D=8$ and $D=9$ dimensional pure Lovelock BHs, respectively, in their maximally rotating configurations. The left panels show the results for different quantities of the plasma magnetization parameter, $\sigma_0\in\{10,10^2,10^3,10^4,10^5\}$, with the orientation angle fixed to $\xi=\pi/20$. The right panels present the connections for varying values of the orientation angle, $\xi\in\{0,\pi/20,\pi/12,\pi/8,\pi/6\}$, with the magnetization parameter fixed to $\sigma_0=10^4$. As shown in Figs.~\ref{fig:6D7DPower} and~\ref{fig:8D9DPower}, the power extracted from the BHs through the Comisso--Asenjo mechanism increases monotonically with higher values of the plasma magnetization parameter and smaller values of the reconnecting magnetic-field orientation angle for all $D=6,\,7,\,8,\,9$ dimensional BHs. However, the BH in $D=9$ dimension shows a more complex behavior than the other cases, while the extracted power reaches considerably higher values, therefore this suggests that energy extraction may be more efficient in nine dimensions. The vertical dashed lines in all figures represent limiting circular orbits. As the magnetic reconnection location approaches the limiting circular orbit, the extracted power decreases dramatically. It is worth noting that, in $D=7$ dimensional pure Lovelock BH, the connections between the extracted power and reconnection location are very similar to the case of $D=4$ dimensional rotating Kerr BH in Einstein gravity.

We now analyze energy extraction efficiencies of the Comisso-Asenjo MR mechanism for each $D=6,\,7,\,8,\,9$ dimensional BHs. To assess the efficiency of the mechanism, the quantities of the accelerated and decelerated plasma energies are very important. In the mechanism, plasma outflows extract the BHs energy using magnetic reconnection process and go to infinity, and plasmas with negative energy fall into the BHs. Therefore the energy extraction efficiency can be written as 
 \begin{eqnarray}
    \eta=\frac{\epsilon_{+}^{\infty}}{\epsilon_{+}^{\infty}+\epsilon_{-}^{\infty}}\, ,
\end{eqnarray}
where $\epsilon_{+}$ and $\epsilon_{-}$ refer to the accelerated and decelerated plasma energies at infinity per unit enthalpy as defined in Eq.~(\ref{Eq:essential}). The energy extraction from the BHs can be done when $\eta>1$ condition satisfy.

In Fig.~\ref{fig:efficiency} energy extraction efficiency of the Comisso--Asenjo mechanism as a function of radial magnetic reconnection locations described in fixed quantities of the magnetization parameter $\sigma_0=100$ and the orientation angle $\xi=\pi/20$ for $D=6$ (upper-left panel), $D=7$ (upper-right panel), $D=8$ (lower-left panel) and $D=9$ (lower-right panel) dimensional pure Lovelock BHs. To examine the maximum energy-extraction rate for each BH, we choose rotation parameters close to their corresponding maximum values. It is clear from Fig.~\ref{fig:efficiency} that, for the $D=6$ dimensional BH with a single rotation, the energy-extraction efficiency can reach nearly $135\%$ in the maximally rotating case. It is worth noting that, in seven dimensions, the energy-extraction efficiency is comparable to that of a $D=4$ dimensional rotating Kerr BH in Einstein gravity, reaching approximately $150\%$. In the $D=8$ dimensional BH case, the efficiency rate exceeds 100$\%$ and even reaches 180$\%$. However, as mentioned above, the power extracted from the $D=9$ dimensional BH is considerably higher than the other cases, i.e. its energy-extraction efficiency can reach its highest values, as shown in Fig.~\ref{fig:efficiency}. The causes for a significantly high energy extraction efficiency for the $D=9$ dimensional BH arise from the strong influence of spacetime geometry and the large spin value that causes frame-dragging effects. This also leads to the formation of negative-energy plasma components in the near-horizon regions.

Another remarkable mechanism for extracting rotational energy from a BH by using magnetic fields and frame-dragging effects is the Blandford--Znajek mechanism\cite{Blandford1977}. The energy extraction power can be extracted via the Blandford--Znajek mechanism under maximum efficiency conditions and can be expressed as~\cite{Tchekhovskoy10ApJ}
\begin{eqnarray}
   P_{BZ} = \kappa \Phi_{BH}^2 \Omega_{H}^2(1+\chi_1\Omega_{H}^2+\chi_2\Omega_{H}^4+{\cal O}({\Omega_{H}^6}))\, ,
\end{eqnarray}
where $\Phi_{\text{BH}}$ is the magnetic flux and $\Omega_{H}$ is the angular frequency of BH horizon while $\kappa$, $\chi_1$, $\chi_2$ are treated as numerical constants determined by the BH geometry and the properties of the magnetic field. In our calculations, assuming a split monopole geometry, we adopt $\kappa=0.05$, $\chi_1=1.38$, and $\chi_2=-9.2$~\cite{Tchekhovskoy10ApJ}. To compare BH extracted power quantities with the fast occurring Comisso-Asenjo MR mechanism, the magnetic flux can be simplified as $\Phi_{\text{BH}}\sim B_0 \sin \xi \, r_H^2$ and the comparison can be written as
 \begin{eqnarray}
\frac{P_{ext}}{P_{BZ}} = \frac{-\epsilon_{-}^{\infty} A_{in} U_{in}}{\kappa \Omega_{H}^2r_{H}^4\sigma_0\sin^2{\xi}(1+\chi_1\Omega_{H}^2+\chi_2\Omega_{H}^4)}\, .
\end{eqnarray}

Figure~\ref{fig:PowerMRBZ} compares the power extracted from the BHs via the Comisso--Asenjo and Blandford--Znajek mechanisms as a function of the magnetization parameter for the fast non-collisional reconnection regime when the orientation angle is fixed at $\xi=\pi/20$ for maximally rotating pure Lovelock BHs with a single rotation in $D=6$ (upper-left panel), $D=7$ (upper-right panel), $D=8$ (lower-left panel), and $D=9$ (lower-right panel) dimensions. Comparisons are made for different radii of magnetic reconnection locations. Note that $P_{MR}/P_{BZ}\gg1$ satisfies for an extended range of magnetization values, as exhibited in Fig.~\ref{fig:PowerMRBZ}. As a result, the Comisso-Asenjo MR mechanism that is more efficient than the Blandford-Znajek process in these regions of all higher dimensional pure Lovelock BHs. In the case in which $\sigma_0\to\infty$, the power extracted through the Blandford-Znajek mechanism is always much higher than MR mechanism as that of magnetization limit, $P_{ext}/P_{BZ}\to0$. It can be seen from Fig.~\ref{fig:PowerMRBZ} that the ratio is considerably higher for the $D=9$ dimensional BH than for other dimensional cases.

\section{Discussion and conclusion}
\label{Sec:con}

The extraction of rotational energy from BHs is one of the most significant predictions of general relativity. Consequently, understanding the energetics of rotating BHs is essential for explaining high-energy astrophysical phenomena. With this in mind, in this paper, we extended the Comisso--Asenjo MR mechanism to rotating pure Lovelock/GB BHs with a single rotation parameter. Restricting our analysis to dimensions $2N+2\leq D\leq 4N+1$, we performed an investigation of the efficiency and power of rotational energy extraction, examining their dependence on BH spin, plasma magnetization, magnetic field orientation, and the location of the reconnection region. Our study provides new insights into the Comisso--Asenjo MR mechanism in higher-dimensional rotating BH spacetimes, revealing distinctive features of energy extraction in pure Lovelock gravity. Our theoretical findings provide a basis for future observational and experimental studies of energy extraction in higher dimensional gravity.

We examined the accelerated and decelerated plasma energies to assess energy extraction mechanism and its efficiency. By analyzing the dependence of the accelerated and decelerated plasma energies on the magnetization parameter for $D=6,\,7,\,8,\,9$ dimensional BHs, we concluded that both energy components are strongly influenced by the magnetization parameter. In particular, the accelerated plasma energy increases monotonically, whereas the decelerated plasma energy decreases monotonically as the magnetization parameter grows. Then, we investigated parameter-space regions where the energy extraction through Comisso-Asenjo MR mechanism can be applied where $\epsilon^{\infty}_{+}>0$ and $\epsilon^{\infty}_{-}<0$ for varying quantities of plasma magnetization parameter and orientation angle in all allowed dimensions considered in this study. For all higher dimensional cases, the parameter space regions supporting energy extraction expand with increasing the plasma magnetization and decreasing magnetic field orientation angle, indicating that highly magnetized plasmas with the corresponding field orientations enhance the MR process. This indicates that higher plasma magnetization and smaller magnetic field orientation angles favor more efficient energy extraction through the Comisso--Asenjo MR mechanism.

We further analyzed the extracted power via the Comisso-Asenjo MR mechanism from maximally rotating pure Lovelock BHs having a single rotation in dimensions $D=6,\,7,\,8,\,9$  as a function of the reconnection location radius for different values of the plasma magnetization and magnetic field orientation angle. For all dimensions considered, larger plasma magnetization and smaller magnetic field orientation angles lead to a significant enhancement of the extracted power. Although the overall behavior is qualitatively similar in $D=6,\,7,\,8,\,9$, the extracted power attains a substantially higher peak power in dimension $D=9$ than in lower-dimensional cases. We also examined energy extraction efficiency from higher-dimensional pure Lovelock BHs through the Comisso-Asenjo MR mechanism. Similarly, as the spacetime dimension increases, the efficiency of energy extraction also increases in each dimension, and is further enhanced as the spin parameter approaches its maximal value. In particular, we found that the highest energy extraction efficiency is achieved in the $D=9$ dimensional case. This behavior may arise from stronger frame-dragging effects and modifications of the horizon and ergosphere structures in higher dimensional spacetime geometry, which facilitate the formation of negative-energy plasma components.

Finally, we compared the power extracted from higher-dimensional pure Lovelock BHs through two distinct energy extraction mechanisms, i.e. the Comisso--Asenjo MR mechanism and the Blandford--Znajek process. In the latter, rotational energy is extracted electromagnetically as frame dragging twists the magnetic field lines threading the event horizon. Our results showed that, over a broad range of magnetization parameter values, the Comisso--Asenjo MR mechanism is substantially more efficient than the Blandford--Znajek process, with $P_{MR}/P_{BZ}\gg1$ in all higher-dimensional cases considered. However, in the limit of an extremely large magnetization parameter $(\sigma_{0}\to\infty)$, the Blandford--Znajek process becomes the dominant energy extraction mechanism, as $(P_{\rm ext}/P_{\rm BZ}\to 0)$. Notably, we showed that the ratio $(P_{\rm MR}/P_{\rm BZ})$ is significantly larger for the $D=9$ dimensional pure Lovelock BH case. 

Interestingly, we found that the $D=7$ dimensional pure Lovelock BH exhibits the same behavior as its four-dimensional Einstein counterpart. In these dimensions, the two theories coincide, leading to remarkably similar spacetime structures. Consequently, the Comisso--Asenjo MR mechanism exhibits essentially the same energy extraction efficiencies and extracted powers in both cases.

\appendix

\bibliographystyle{apsrev4-1}  
\bibliography{gravreferences,Ref_BS1,Ref_BS2}

\end{document}